\documentclass[aps,prd,superscriptaddress,amsmath,groupedaddress,twocolumn,showpacs,nofootinbib]{revtex4-1}

\usepackage[pdftex]{color}
\usepackage[dvipsnames]{xcolor}
\usepackage[sort&compress]{natbib}
\usepackage{color,hhline,psfrag,rotating,amssymb}
\usepackage[colorlinks=true,linkcolor=PineGreen,filecolor=PineGreen,urlcolor=PineGreen,citecolor=PineGreen,pdftex,plainpages=false]{hyperref}
\usepackage[hang,nooneline]{subfigure}
\usepackage{dcolumn}

\usepackage{graphicx}
\usepackage{revsymb}
\usepackage{dcolumn}
\newcommand{\msb}{\mathrm{\overline{MS}}}

\newcommand{\lat}{\mathrm{lat}}

\newcommand{\datadiff}{-0.69(23)}
\newcommand{\mbmcetah}{4.578(12)}
\newcommand{\etahRmbmc}{1.000372(90)}
\newcommand{\mbmcQetah}{4.586(13)}
\newcommand{\mbmcpsih}{4.578(15)}
\newcommand{\psihRmbmc}{1.00036(19)}
\newcommand{\mbmcQpsih}{4.586(15)}
\newcommand{\mbmcetahpsih}{4.578(12)}
\newcommand{\mbpsietadiff}{0.01(23)}

\newcommand{\mbmcetahQpsihQ}{4.586(12)}
\newcommand{\mbetahQpsihQ}{4.513(26)}
\newcommand{\mcetahQpsihQ}{0.9841(51)}
\newcommand{\mbmfiveetahQpsihQ}{4.202(21)}
\newcommand{\mbmfouretahQpsihQ}{4.209(21)}
\newcommand{\metac}{2.9766(13)}
\newcommand{\metab}{9.3987(22)}
\newcommand{\mpsi}{3.09620(20)}
\newcommand{\mups}{9.46030(26)}
\newcommand{\Rmc}{0.99823(17)}
\newcommand{\Rmcred}{0.99867(13)}

\newcommand{\red}[1]{{\color{red} #1}}
\renewcommand{\red}[1]{{ #1}}     

\usepackage{times}

\begin{document}
\title{Determination of $\overline{m}_b/\overline{m}_c$ and $\overline{m}_b$ from 
$n_f=4$ lattice QCD\,+\,QED}
        \author{D.~Hatton}
        \affiliation{SUPA, School of Physics and Astronomy, University of Glasgow, Glasgow, G12 8QQ, UK}
        \author{C.~T.~H.~Davies}
        \email[]{christine.davies@glasgow.ac.uk}
        \affiliation{SUPA, School of Physics and Astronomy, University of Glasgow, Glasgow, G12 8QQ, UK}
        \author{J.~Koponen}
        \affiliation{PRISMA+ Cluster of Excellence and Institute for Nuclear Physics,
        Johannes Gutenberg University of Mainz,
        D-55128 Mainz, Germany}
        \author{G.~P.~Lepage}
        \email[]{g.p.lepage@cornell.edu}
        \affiliation{Laboratory for Elementary-Particle Physics, Cornell University, Ithaca, New York 14853, USA}
        \author{A.~T.~Lytle}
        \affiliation{Department of Physics, University of Illinois, Urbana, IL 61801, USA}
        \collaboration{HPQCD collaboration}
        \homepage{http://www.physics.gla.ac.uk/HPQCD}
        \noaffiliation
\date{\today}

\begin{abstract}
We extend HPQCD's earlier \red{$n_f=2+1+1$} lattice-QCD analysis of the ratio of $\msb$~masses of the $b$~and $c$~quark to include results from finer lattices (down to 0.03\,fm) and a new calculation of QED contributions to the mass ratio. We find that $\overline{m}_b(\mu)/\overline{m}_c(\mu)=\mbmcetahQpsihQ$ at renormalization scale~$\mu=3$\,GeV. This result is nonperturbative. Combining it with HPQCD's recent lattice QCD\,+\,QED determination of~$\overline{m}_c(3\,\mathrm{GeV})$ gives a new value for the $b$-quark mass: $\overline{m}_b(3\,\mathrm{GeV}) = \mbetahQpsihQ$\,GeV. The $b$-mass corresponds to $\overline{m}_b(\overline{m}_b, n_f=5) = \mbmfiveetahQpsihQ$\,GeV. These results are the first based on simulations that include QED. 
\end{abstract}

\maketitle
\section{Introduction}
Accurate masses for heavy quarks are important for QCD-phenomenology generally, but they will be particularly important for high-precision searches for new physics in Higgs decays~\cite{Lepage:2014fla}. In this paper we present a new result for the ratio~$\overline{m}_b/\overline{m}_c$ of the $\msb$~masses of the $b$~and $c$~quarks. Our analysis of the mass ratio is completely nonperturbative. This is in contrast to lattice-QCD determinations of the separate quark masses, which need QCD perturbation theory to relate $\msb$~masses to lattice quantities. Thus the (nonperturbative) mass ratio provides a nontrivial check on  (perturbative) determinations of the separate masses. The ratio can also be combined with recent accurate determinations of the $c$-quark mass to obtain new results for the $b$-quark mass.    

Lattice simulations of $b$~quarks are complicated by the quark's large mass, which leads to large lattice-spacing errors when the $b$~quarks are described by the Dirac equation (as opposed to, say, NRQCD~\cite{Lee:2013mla}). We address this problem by using a Highly Improved Staggered-Quark discretization of the Dirac equation (HISQ~\cite{Follana:2006rc}) that is also highly efficient, making simulations at very small lattice spacings feasible. Our previous analysis of the mass ratio~\cite{Chakraborty:2014aca} used lattices with spacings down to~0.06\,fm but still required an extrapolation in the quark mass to reach~$m_b$. Here we reduce the lattice spacing to~0.03\,fm, where $am_b\approx0.6$, which allows us to simulate at the $b$~mass. Lattice spacing errors at~$m_b$ are less than~1\% on our finest lattice, and we are able to remove most of that error by extrapolating from results covering a range of lattice spacings and heavy-quark masses~$m_h$.

Our new result is accurate to about~0.25\%, so it becomes important to include QED effects. We recently analyzed the QED contributions to the $c$~quark's mass~\cite{Hatton:2020qhk}. Here we adapt the methods from our earlier paper to provide the first results for QED~contributions to $\overline{m}_b/\overline{m}_c$ and~$\overline{m}_b$. Here and in our earlier paper we use the \textit{quenched} QED approximation, which omits contributions from photons coupling to sea quarks. The quenched approximation should capture the bulk of the QED correction in mesons whose valence quarks are both heavy; contributions from sea quarks are expected to be an order of magnitude smaller~\cite{Hatton:2020qhk}.

In Section~\ref{sec:simulations} we describe our general strategy and the lattice QCD simulations we employed. In Section~\ref{sec:withoutQED}, we extract a value for $\overline{m}_b/\overline{m}_c$ using results from simulations without QED. We then add QED effects in Section~\ref{sec:withQED}. We summarize our results for the mass ratio in Section~\ref{sec:conclusions} and combine them with HPQCD's recent $c$-quark mass to obtain a new result for the $b$-quark mass.

\begin{table*}
    \caption{Gluon configuration sets used in this paper. Sets 
    are grouped by approximate lattice spacing, with lattice
    spacings of 0.09\,fm (Sets~1 and~2), 0.06\,fm (Sets~3 and~4),
    0.045\,fm (Set~5), and 0.03\,fm (Set~6). Lattice spacings are determined from the values shown for the  Wilson flow parameter~$w_0/a$~\cite{Borsanyi:2012zs} where $w_0=0.1715(9)$\,fm~\cite{Dowdall:2013rya}. The sea quark masses are given in lattice units for $u/d$ quarks ($a m_\ell$), $s$~quarks ($a m_s$), and $c$~quarks ($a m_c$). Tuned values for the lattice $c$~masses are also given (in GeV). These masses are adjusted to give correct masses for either the~$\eta_c$ or $J/\psi$~mesons (Eq.~(\ref{eq:mcc_cont})). The $c$~masses are tuned
    using slope~$d\tilde m_c/dm_{cc}$, which is the same (within errors)
    for the $\eta_c$ and $J/\psi$. The spatial and temporal sizes
    of the lattices, $L$ and $T$, are listed, \red{as are the number of 
    configurations used in our analysis (the two numbers for Set~1 are 
    for quark masses $am_h$ below and above~0.5; the three numbers 
    for Set~2 are for the pseudoscalar correlators, the vectors with mass 
    with mass below~0.5 and the vectors with mass above~0.5). 
    The three polarizations were averaged for vectors. Eight~time sources
    were used on each configuration except for Set~6 where four were used.}}
    \label{tab:config}
    \begin{ruledtabular}
    \begin{tabular}{clllllllccc}
    Set & $w_0/a$ & $am_\ell^\mathrm{sea}$ & $am_s^\mathrm{sea}$ & $am_c^\mathrm{sea}$ & $\tilde m_c^\mathrm{tuned}(\eta_c)$ & $\tilde m_c^\mathrm{tuned}(J/\psi)$ & $d\tilde m_c/dm_{cc}$ & $L/a$ & $T/a$ & $N_\mathrm{cfg}$\\
    \hline
    \hline
    1 & 1.9006(20) & 0.0074 & 0.037 & 0.44 & 0.9767 (25) & 0.9828 (27) & 0.478 (10) & 32 & 96 & 300, 504\\
    2 & 1.95175(70) & 0.0012 & 0.0363 & 0.432 & 0.9671 (25) & 0.9717 (27) & 0.478 (10) & 64 & 96& 311, 565, 792\\
    \hline
    3 & 2.8960(60) & 0.0048 & 0.024 & 0.286 & 0.9078 (24) & 0.9118 (26) & 0.444 (10) & 48 & 144& 333\\
    4 & 3.0170(23) & 0.0008 & 0.022 & 0.26 & 0.8944 (23) & 0.8966 (25) & 0.444 (10) & 96 & 192 & 100\\
    \hline
    5 & 3.892(12) & 0.00316 & 0.0158 & 0.188 & 0.8646 (26) & 0.8675 (29) & 0.433 (10) & 64 & 192 & 200\\
    \hline
    6 & 5.243(16) & 0.00223 & 0.01115 & 0.1316 & 0.8234 (27) & 0.8251 (29) & 0.423 (10) & 96 & 288 & 100\\
    \end{tabular}
    \end{ruledtabular}
\end{table*}

\section{Lattice QCD Simulations}
\label{sec:simulations}
We use gluon configuration sets generated on a variety of lattices (by the MILC collaboration~\cite{Bazavov:2012xda}), with $n_f=4$ flavors of HISQ sea quark and lattice spacings ranging from 0.09\,fm to~0.03\,fm. These sets are described in Table~\ref{tab:config}. The $u$~and $d$~quark masses are set equal to $m_\ell\equiv(m_u+m_d)/2$; corrections  to this approximation are quadratic in the light-quark masses for our analysis, and so are negligible. We include results where the light-quark masses in the sea are tuned close to their physical value, but we also include results with much larger light-quark masses in the sea. Results from these last simulations are unphysical but are easily corrected~\cite{Chakraborty:2014aca}. Increasing the light-quark mass in the sea significantly reduces the cost of our analysis at small lattice spacings (because smaller lattice volumes are used).

Ignoring QED for the moment, the ratio of the~$b$ and~$c$ $\msb$~masses equals the ratio of the corresponding bare quark masses used in the lattice Lagrangian, up to corrections that vanish in the continuum limit~\cite{[\red{Ratios of bare masses are scheme independent in pure QCD. For more discussion see: }]Davies:2009ih}:
\begin{equation}
    \frac{\overline{m}_b(\mu)}{\overline{m}_c(\mu)} = \frac{m_{b}^\mathrm{tuned}}{m_{c}^\mathrm{tuned}}\Bigg|_\mathrm{latt}
    + \mathcal{O}(\alpha_s(\pi/a) a^2)
    \label{eq:nonpert_mbmc}
\end{equation} 
where the bare masses are tuned so that the QCD simulations reproduce the experimental results for meson masses. This relationship between the $\msb$ and lattice quark masses is nonperturbative and independent of the $\msb$~renormalization scale~$\mu$.

\begin{figure}
    \includegraphics[scale=0.9]{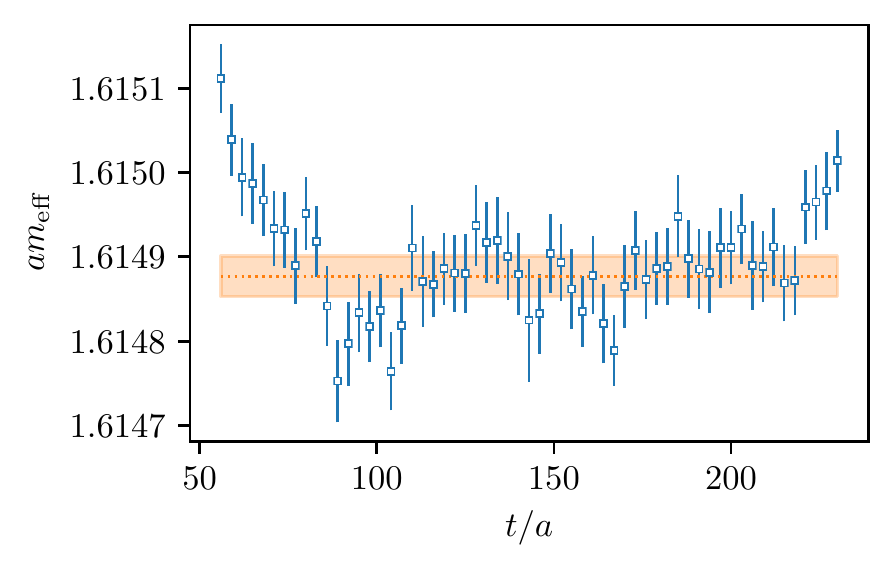}
    \caption{\label{fig:meff} \red{The effective mass plotted versus 
    time for the $am_h=0.65$ pseudoscalar correlator 
    from configuration Set~6. For clarity, the plot includes only 
    every third point. The orange band and dotted line show the corresponding mass (Table~\ref{tab:amhh}) 
    obtained from a multi-exponential fit~\cite{Hatton:2020qhk, Hatton:2021dvg}. The error in the fit result (orange band) is 
    almost entirely statistical in origin. In  particular, 
    possible biases due to excited states 
    are completely negligible ($50\times$~smaller),  as 
    is typical in fits for heavy-quark ground-state masses.
    }}
\end{figure}

Pseudoscalar and vector meson masses from our simulations are listed in Table~\ref{tab:amhh} for a variety of (valence) heavy-quark masses, ranging on the finest lattices (Sets~5 and~6) approximately from the $c$~mass to the $b$~mass. The analysis methods for extracting these masses (and most of the results) come from~\cite{Hatton:2020qhk, Hatton:2021dvg}. \red{We use multi-exponential fits to calculate the masses. Fig.~\ref{fig:meff}
compares the result from our fit with the effective mass values 
at various times for the correlator closest to the $\eta_b$ mass on our finest lattice.}

The quark masses $am_h$ in Table~\ref{tab:amhh} are what is used in the HISQ Lagrangian. The $a\tilde m_h$ masses are corrected  to remove tree-level \red{$(am_h)^{2n}$~errors (in the pole mass) through order $2n=10$}~\cite{Follana:2006rc, McLean:2019sds}:
\begin{align}
a\tilde{m}_h &\equiv a m_h \Bigg(
    1 
    - \frac{48}{80}\Big(\frac{am_h}{2}\Big)^4
    + \frac{1472}{2240}\Big(\frac{am_h}{2}\Big)^6\nonumber \\
    &+ \frac{456448}{537600}\Big(\frac{am_h}{2}\Big)^8
    - \frac{78789632}{23654400}\Big(\frac{am_h}{2}\Big)^{10}
    \Bigg).
    \label{eq:amhtilde}
\end{align}
We write the expansion as powers of $am_h/2$ because this makes the leading coefficients roughly the same size (about~$1/2$). The correction is~$-2$\% at $am_h=0.9$, which is the largest value we use.

We give results in Table~\ref{tab:config} for the tuned bare $c$~mass for each of the configurations. In each case we adjust the $c$~mass so as to reproduce the continuum value for either the $\eta_c$~mass or the $J/\psi$~mass:
\begin{align}
    m_{\eta_c}^\mathrm{cont}=\metac\,\mathrm{GeV} \nonumber \\
    m_{J/\psi}^\mathrm{cont}=\mpsi\,\mathrm{GeV}.
    \label{eq:mcc_cont}
\end{align}
Here we have subtracted 7.3(1.2)\,MeV from the experimental value for $m_{\eta_c}$~\cite{Zyla:2020zbs} to account for the fact that we are not including contributions from $c\overline{c}$~annihilation in our simulations; this correction is determined in~\cite{Hatton:2020qhk}. The analogous correction to the $J/\psi$~mass is negligible, but we have subtracted 0.7(2)\,MeV from the mass to account for $c\overline{c}$~annihilation to a photon; this correction is estimated perturbatively in~\cite{Hatton:2020qhk}. We extrapolate the $c$~masses to their correct values using
\begin{equation}
    a\tilde m_c^\mathrm{tuned} = a\tilde m_c - \big(am_{cc}^\lat - a m_{cc}^\mathrm{cont}\big) \frac{d\tilde m_c}{dm_{cc}}
\end{equation}
where $m_{cc}$ is either the $\eta_c$ or the $J/\psi$~mass and the slopes (see Table~\ref{tab:config}) are estimated from splines fit to the entries in Table~\ref{tab:amhh}. 

\begin{table}
    \caption{Lattice QCD results for the ground-state 
    pseudoscalar and vector $h\overline{h}$~mesons in lattice units:
    $a m_{hh}^P$ and $a m_{hh}^V$, respectively. Results 
    are given for each configuration sets (Table~\ref{tab:config}) and a variety of bare quark masses $am_h$ and corrected masses $a\tilde m_h$ (in lattice 
    units). The uncertainties in the meson masses are negligible compared with
    other errors in our analysis and so have no impact on the final results.
    \red{Most of these results are from~\cite{Hatton:2020qhk, Hatton:2021dvg}.}}
    \label{tab:amhh}
    \begin{ruledtabular}
    \begin{tabular}{ccccc}
    Set & $am_h$ & $a \tilde{m}_h$ & $a m_{hh}^P$ & $a m_{hh}^V$\\
    \hline
    1& 0.45 & 0.44935 & 1.366803 (89) & 1.41567 (21)\\
    & 0.6 & 0.59739 & 1.675554 (47) & 1.717437 (70)\\
    & 0.8 & 0.79003 & 2.064088 (40) & 2.101542 (57)\\
    2& 0.433 & 0.43246 & 1.329290 (31) & 1.378280 (54)\\
    & 0.6 & 0.59739 & 1.674264 (13) & 1.715453 (32)\\
    & 0.8 & 0.79003 & 2.063015 (11) & 2.099940 (26)\\
    \hline
    3& 0.269 & 0.26895 & 0.885242 (56) & -- \\
    & 0.274 & 0.27394 & 0.896664 (33) & 0.929876 (86)\\
    & 0.4 & 0.39963 & 1.175559 (29) & 1.202336 (85)\\
    & 0.5 & 0.49891 & 1.387459 (27) & 1.411113 (72)\\
    & 0.6 & 0.59739 & 1.593089 (25) & 1.614626 (63)\\
    & 0.7 & 0.69461 & 1.793118 (23) & 1.813249 (57)\\
    & 0.8 & 0.79003 & 1.987504 (22) & 2.006783 (52)\\
    4& 0.26 & 0.25996 & 0.862671 (27) & 0.895702 (52)\\
    & 0.4 & 0.39963 & 1.173904 (23) & 1.199806 (36)\\
    & 0.6 & 0.59739 & 1.591669 (19) & 1.612586 (27)\\
    & 0.8 & 0.79003 & 1.986246 (17) & 2.005047 (24)\\
    \hline
    5& 0.194 & 0.19399 & 0.666821 (41) & 0.692026 (59)\\
    & 0.4 & 0.39963 & 1.130722 (31) & 1.147617 (40)\\
    & 0.6 & 0.59739 & 1.549098 (26) & 1.562884 (32)\\
    & 0.8 & 0.79003 & 1.945787 (23) & 1.958252 (27)\\
    & 0.9 & 0.88303 & 2.135642 (21) & 2.147903 (25)\\
    \hline
    6& 0.138 & 0.13800 & 0.496969 (42) & 0.516149 (61)\\
    & 0.45 & 0.44935 & 1.201328 (29) & 1.211601 (28)\\
    & 0.55 & 0.54828 & 1.410659 (27) & 1.420048 (24)\\
    & 0.65 & 0.64619 & 1.614877 (24) & 1.623684 (21)\\
    \end{tabular}
    \end{ruledtabular}
\end{table}

\section{$\overline{m}_b/\overline{m}_c$ without QED} 
\label{sec:withoutQED}
We  could tune the lattice $b$~mass the same way we tuned~$m_c$, but we have only a few simulation results near the~$b$ and these have significant $(am_h)^{2n}$ errors. Instead we will use the data in Table~\ref{tab:amhh} to define functions that relate the ratio of quark masses to the pseudoscalar ($P$) or vector ($V$)~masses:
\begin{align}
    m_{hh}^{P} &= f_{hh}^{P}(\overline{m}_h/\overline{m}_c) \nonumber \\
    m_{hh}^{V} &= f_{hh}^{V}(\overline{m}_h/\overline{m}_c),
    \label{eq:fhh}
\end{align}
where 
\begin{align}
    f_{hh}^{P}(1) &\equiv m_{\eta_c}^\mathrm{cont} \nonumber \\ 
    f_{hh}^{V}(1) &\equiv m_{J/\psi}^\mathrm{cont}.
\end{align}
Given these functions, we then obtain two estimates for $\overline{m}_b/\overline{m}_c$ by solving each of the equations
\begin{align}
    f_{hh}^{P}(\overline{m}_b/\overline{m}_c) &= m_{\eta_b}^\mathrm{cont} \nonumber \\ 
    f_{hh}^{V}(\overline{m}_b/\overline{m}_c) &= m_{\Upsilon}^\mathrm{cont}
    \label{eq:solve_fhh}
\end{align}
for $\overline{m}_b/\overline{m}_c$, where 
\begin{align}
    m_{\eta_b}^\mathrm{cont} &= \metab\,\mathrm{GeV} \nonumber \\
    m_{\Upsilon}^\mathrm{cont} &= \mups\,\mathrm{GeV}.
    \label{eq:mbb_cont}
\end{align}
The two mass ratios should agree.
Here we account for the $b\overline{b}$~annihilation contribution to the ${\eta_b}$~mass by adding an extra error of $\pm1$\,MeV to the experimental result~\cite{Zyla:2020zbs}; this estimate is based on NRQCD perturbation theory and the meson's width~\cite{Follana:2006rc}. The analogous contribution to the $\Upsilon$~mass is negligible, as is $\Upsilon$ annihilation via a photon.

In what follows, we first describe our lattice-QCD analysis of $f_{hh}^P$ and~$f_{hh}^V$, and then discuss the results.

\subsection{Analysis}
We determine the $f_{hh}$ functions (Eq.~(\ref{eq:fhh})) by fitting the meson masses~$am_{hh}$ from Table~\ref{tab:amhh} to functions of the following form,
\begin{align}
    am_{hh} \times (1 \pm \sigma_u) &= a f_{hh}(r) \nonumber \\
    &+ am_{hh} \big(\delta_{a^2} + \delta_{uds}^\mathrm{sea} + \delta_c^\mathrm{sea} \big),
\label{eq:model}
\end{align}
where 
\begin{equation}
    r \equiv 
    \frac{a\tilde m_h}{a\tilde m_c^\mathrm{tuned}}\,
    \frac{\xi_m(\tilde m_c^\mathrm{tuned}, \delta m_{uds}^\mathrm{sea}) }{\xi_m(\tilde m_h, \delta m_{uds}^\mathrm{sea})}
    \label{eq:r}
\end{equation}
is the effective ratio of quark masses~$\overline{m}_h/\overline{m}_c$. Pseudoscalar and vector mesons are fit separately, to determine each of $f^P_{hh}$ and~$f^V_{hh}$.
We describe each element of the fit function in turn:
\begin{itemize}
    \item We increase the fractional error on each value of~$am_{hh}$ 
    from Table~\ref{tab:amhh} to $\pm\sigma_u$. The $am_{hh}$~errors listed in the table are very small. It is impossible to fit the almost six~significant digits in these data with a model as simple as we use here. So we increase the fractional error on each value to~$\sigma_u$, which is then a measure of the part of the variation in the data that is \textit{unexplained} by our model. 
    The $\sigma_u$~errors are uncorrelated from one $am_{hh}$ to another.
    We use the same value for $\sigma_u$ for every data point and adjust its size to maximize the Bayes Factor from the fit~\cite{[See Section 5.2 on the Empirical Bayes criterion in ]Lepage:2001ym}. For the parameters and model used here, we find that 
    \begin{equation}
        \sigma_u = 0.00025,
    \end{equation}
    which means that our model explains the individual data points to within $\pm0.025$\%. A simpler model would have a larger~$\sigma_u$: for example, $\sigma_u$~more than doubles if the~$\xi_m$ factors in ratio~$r$ are dropped \red{(but gives consistent results within the larger errors)}. Note that the statistical errors listed in~Table~\ref{tab:amhh} can be neglected when $\sigma_u$~is included.

    \item We parameterize the $f_{hh}$ functions as splines~\cite{[We use the monotonic spline described in ]Steffen:1990} with 
    6~knots evenly spaced from~$r=1$ to~$4.6$ ($\approx m_b/m_c$), inclusive. The fit parameters are the function values at the knots.  These functions are linear up to corrections of order $v^2/c^2\sim0.1$--\,0.3, where $v$~is the typical velocity of the heavy quarks in the meson. Therefore we use the following priors for the values at the knots with $r>1$:
    \begin{align}
        f_{hh}(r_\mathrm{knot}) &=
         1.0(2)\times \nonumber \\
        &\Big(m_{cc} + \frac{r_\mathrm{knot} - 1}{3.6} (m_{bb} - m_{cc})
            \Big),
    \end{align}
    where $m_{cc}$ and $m_{bb}$ are the continuum masses of the 
    pseudoscalar/vector mesons composed of~$c$ and $b$~quarks, respectively (Eqs.~(\ref{eq:mcc_cont}) and~(\ref{eq:mbb_cont})). At $r=1$, we require
    \begin{equation}
        f_{hh}(r=1) = m_{cc}.
    \end{equation}
    We choose 6~knots to maximize the Bayes Factor from the fit. 
    Results obtained using~5 or~7 knots agree well with those from 6~knots, \red{with similar or smaller errors.} Doubling the width of the priors has no effect on our results.

    \item The $\xi_m$~factors in the mass ratio~$r$ rescale the quark masses to correct for detuned values of the light sea quarks. From~\cite{Chakraborty:2014aca},
    \begin{equation}
        \xi_m(m_h, \delta m_{uds}^\mathrm{sea}) = 1 + \frac{g_m}{(m_h/m_c)^\zeta} \,\frac{\delta m_{uds}^\mathrm{sea}}{m_s}
    \label{eq:xim}
    \end{equation}
    where 
    \begin{align}
        \delta m_{uds}^\mathrm{sea} &\equiv \sum_{q=u,d,s}\big(m_q^\mathrm{sea} - m_q^\mathrm{tuned}\big)
    \end{align}
    is the difference between the masses used in the simulation and their tuned values. The 
    tuned masses are determined from the tuned $c$-quark mass using results for $m_c/m_s$ and $m_s/m_\ell$ from~\cite{Bazavov:2018omf}.
    The priors for the fit parameters are 
    \begin{align} 
        g_m = 0.035(5) \quad\quad
        \zeta = 0.3(1),
    \end{align}
    which come from fits described in~\cite{Chakraborty:2014aca}.

    \item The largest simulation errors are from the discretization. These are suppressed by $\alpha_s(\pi/a)$ in order~$a^2$ because we are using the HISQ formalism~\cite{Follana:2006rc}. Beyond this order they are suppressed either by $\alpha_s(\pi/a)$ or by $v^2/c^2$, since we have removed the tree-level $a^{2n}$~errors in the quark masses using Eq.~(\ref{eq:amhtilde}). The fit can't distinguish easily between $\alpha_s$ and $v^2/c^2$ since both are around~0.2 for our data, so we include only an $\alpha_s$ correction,
    modeled after Eq.~(\ref{eq:amhtilde}):
    \begin{align}
        \delta_{a^2} \equiv \alpha_s(\pi/a) \sum_{n=1}^3
        f_{a^2}^{n}(r) \Bigg(\frac{a\tilde m_h}{2}\Bigg)^{2n}.
    \end{align}
    Here functions $f_{a^2}^{n}(r)$ are 6-knot splines with priors at the knots (same locations as above) of 
    \begin{equation}
        f_{a^2}^n(r_\mathrm{knot}) = 0.0(5).
        \label{eq:discprior}
    \end{equation}
    Terms beyond $n=3$~have no effect on the fit results; \red{keeping just the $n=1$~term gives the same final results but with errors that are 25\%~smaller.} The splines allow for $m_h$ dependence in the $a^{2n}$ corrections.

    \item We include $a^2$~corrections to~$\xi_m$ since $\delta m_{uds}^\mathrm{sea}$
    is large for some of our configuration sets:
    \begin{equation}
        \delta^\mathrm{sea}_{uds} = \alpha_s(\pi/a)\,f_{uds}^\mathrm{sea}(r)\,
        \frac{\delta m_{uds}^\mathrm{sea}}{10 m_s}\,\Bigg(\frac{a\tilde m_h}{2}\Bigg)^2.
    \end{equation}
    where function $f_{uds}^\mathrm{sea}(r)$ is again a 6-knot spline, now with priors 
    at the knots of 
    \begin{equation}
        f_{uds}^\mathrm{sea}(r_\mathrm{knot}) = 0.0(1).
    \end{equation}
    where the width is chosen to be somewhat larger than suggested by~$g_m$ above. \red{Omitting this correction has negligible effect on our 
    final results.}

    \item We also include a correction to~$\xi_m$ from detuned $c$-quark masses in the sea. This correction should be small because of heavy-quark decoupling~\cite{Chakraborty:2014aca}\,---\,the momentum transfers in the heavy-quark mesons are too small to produce $c\overline{c}$~pairs efficiently. We include the correction
    \begin{equation}
        \delta_c^\mathrm{sea} = f_{c}^\mathrm{sea}(r) \, \frac{\delta m_c^\mathrm{sea}}{m_c}
    \end{equation}
    where $\delta m_c\equiv m_c - m_c^\mathrm{tuned}$, and $f_{c}^\mathrm{sea}(r)$ is a 6-knot spline with 
    \begin{equation}
        f_{c}^\mathrm{sea}(r_\mathrm{knot}) = 0.00(1).
    \end{equation}
    We choose the width to maximize the Bayes Factor from the fit.
    \red{Omitting this correction has negligible effect on our 
    final results.}
\end{itemize}
The fit parameters are the values of the coefficient functions (splines) at the knots, together with $g_m$ and~$\zeta$ from the $\xi_m$~factors~(Eq.~(\ref{eq:xim})). We use the \texttt{lsqfit} Python module to do the fits~\cite{peter_lepage_2020_4037174, [The splines are implemented using the \texttt{gvar} Python module: ]peter_lepage_2020_4290884}.

\subsection{Results}
\begin{figure}
    \includegraphics[scale=0.9]{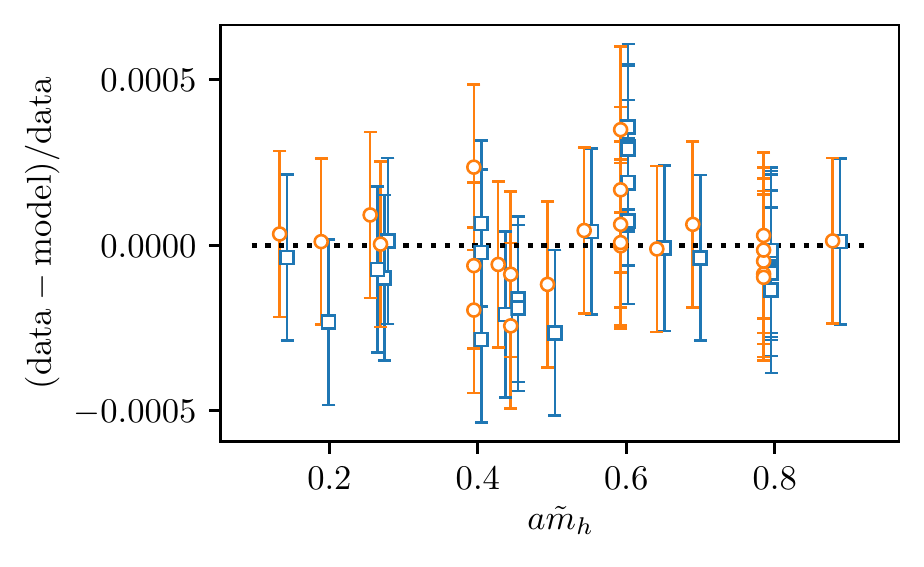}
    \caption{\label{fig:data_model} Relative difference between the data for $am_{hh}$ from Table~\ref{tab:amhh} and the model in Eq.~(\ref{eq:model}) with best-fit values for the fit parameters. Results are shown for both pseudoscalar (squares, offset right) and vector (circles, offset left) mesons, where data points for the two mesons are offset slightly in opposite directions to improve visibility.}
\end{figure}

The functions $f^{P/V}_{hh}$ obtained from the fits described in the previous section (and detailed in the Appendix) are plotted in Fig.~\ref{fig:mhmc_vs_mhh}, together with the data from Table~\ref{tab:amhh}. 
Fig.~\ref{fig:data_model} shows that the model (Eq.~(\ref{eq:model})), with best-fit values for the fit parameters in the corrections (on the right-hand side), reproduces the data within errors.\footnote{$\chi^2$ is less useful as a measure of goodness-of-fit here because we adjust~$\sigma_u$ to give a good fit. $\chi^2$ per degree of freedom was~0.9 for the pseudoscalar data (26~points) and~0.8 for the vector data (25~points).} The difference between the lattice results with and without corrections is~$\datadiff$\% for the highest quark mass on the finest lattice (Set~6).

\begin{figure}
    \includegraphics[scale=0.9]{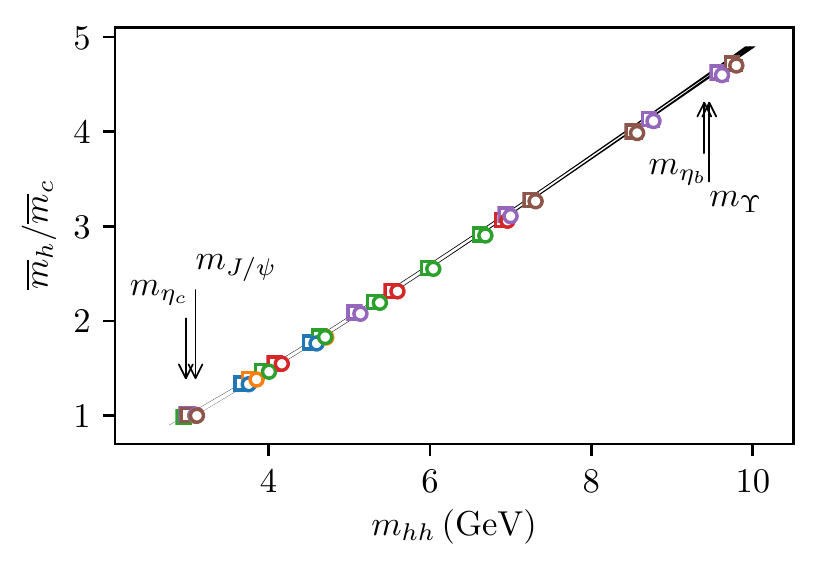}
    \caption{\label{fig:mhmc_vs_mhh} Lattice QCD (without QED) results for $\overline{m}_h/\overline{m}_c$ plotted versus the $h\overline{h}$~meson  masses. Values for $\overline{m}_h/\overline{m}_c$ are corrected as in Eq.~(\ref{eq:r}). The lines, which vary in thickness, show  results from the best-fit values for the functions~$f_{hh}^{P/V}$; the line thickness shows the $1\sigma$~uncertainty in these functions. These functions can be reconstructed from information in the Appendix. Separate results are shown using the pseudoscalar masses $m_{hh}^P$ (top line, squares) and the vector masses $m_{hh}^P$ (bottom line, circles). Different colors indicate different configuration sets, with Sets~6 (brown) and~5 (purple) having the largest masses, followed by Sets~4 (red), 3 (green), 2 (orange) and 1 (blue), in that order. Error bars are smaller than the plot symbols.
    }
\end{figure}

\begin{table}
    \caption{Contributions to the total ($1\sigma$) error in $\overline{m}_b/\overline{m}_c$ from 
    QCD simulations (without QED), as a percentage of the mean value.
    Results are given for determinations using pseudoscalar mesons
    ($m_{hh}^P$) and vector mesons ($m_{hh}^V$), and for the weighted 
    average of these results.
    The dominant errors come from the extrapolation to zero lattice spacing
    and from uncertainties in the lattice spacing. Additional errors
    are from residual uncertainties taken in the fit data ($\sigma_u$), 
    uncertainties in $\xi_m$ used to correct for unphysical sea-quark masses, uncertainties in the~$\eta_c$ and~$\eta_b$ masses, and tuning uncertainties   in the sea-quark masses. The error budgets are the same when QED is included aside from an additional uncertainty 
    of 0.03\% associated with the QED corrections.
    }
    \label{tab:errorbudget}
    \begin{ruledtabular}
    \begin{tabular}{rccc}
     & $\overline{m}_b/\overline{m}_c[m_{hh}^P]$ & $\overline{m}_b/\overline{m}_c[m_{hh}^V]$ & $\overline{m}_b/\overline{m}_c[\mathrm{avg}]$\\
    \hline
    $(am_h)^2\to 0$ & 0.20 & 0.21 & 0.20\\
    $w_0$, $w_0/a$ & 0.10 & 0.18 & 0.12\\
    $\sigma_u$ & 0.12 & 0.12 & 0.09\\
    $g_m$, $\zeta$ & 0.05 & 0.05 & 0.05\\
    $m_{cc}$ & 0.06 & 0.01 & 0.04\\
    $m_{bb}$ & 0.03 & 0.00 & 0.02\\
    $(am_h)^2\delta m_{uds}^\mathrm{sea}\to 0$ & 0.06 & 0.07 & 0.06\\
    $\delta m_c^\mathrm{sea}\to 0$ & 0.03 & 0.03 & 0.03\\
    $d\tilde m_c/dm_{cc}$ & 0.03 & 0.02 & 0.02\\
    \hline
    Total: & 0.27 & 0.32 & 0.27\\
    \end{tabular}
    \end{ruledtabular}
\end{table}

We can use functions $f^{P/V}_{hh}$ to extract values for the ratio of $\msb$~masses by solving Eqs.~(\ref{eq:solve_fhh}). We obtain
\begin{align}
    \overline{m}_b / \overline{m}_c &= 
    \begin{cases}
        \mbmcetah & \mbox{from the $am_{hh}^P$} \\
        \mbmcpsih & \mbox{from the $am_{hh}^V$},
    \end{cases}
\end{align}
independent of renormalization scale.
The two estimates agree to within $\mbpsietadiff$\%. The weighted 
average, taking account of correlations, is 
\begin{align}
    \overline{m}_b / \overline{m}_c = \mbmcetahpsih.
    \label{eq:withoutQED}
\end{align}

We tabulate the leading uncertainties in our two results in Table~\ref{tab:errorbudget}. The error budgets are similar for the two mesons, and are dominated by uncertainties associated with discretization errors and the lattice spacings. \red{Doubling the 
widths of any of the priors associated with these uncertainties 
has negligible effect on the central values  from our fits ($<\sigma/3$), and only doubling the discretization priors (Eq.~(\ref{eq:discprior})) has an appreciable impact on the final uncertainties, as expected from Table~\ref{tab:errorbudget}. Omitting results from the 
coarsest lattices (Sets~1 and~2) has negligible effect on 
our results ($<\sigma/10$). Omitting results from the finest 
lattice (Set~6) increases the final uncertainties significantly (by factors of~5--6) because there is then insufficient data at large masses to constrain the 6-knot splines used in the fit function; 
reducing the number of knots decreases the errors by a third. In either case the results agree with our final results within errors.}

\red{Finally, as discussed in~\cite{Hatton:2020qhk}, we expect errors from the finite lattice volume and strong-isospin breaking ($m_u\ne m_d$) in the sea to be less than~0.01\% and so negligible here. We have verified this for the meson masses (using configuration Sets~3A--3B and~5--7 from~\cite{Hatton:2020qhk}). The Wilson flow parameter~$w_0$ should be similarly insensitive, and we have verified this to the level of our statistical errors~(0.03\%) for~$w_0/a$. See~\cite{Hatton:2020qhk} for further details.
}

\section{Adding QED}
\label{sec:withQED}
Adding QED complicates the analysis of $\overline{m}_b/\overline{m}_c$ because the quarks have different QED charges and therefore different mass anomalous dimensions. Thus the nonperturbative relation in Eq.~(\ref{eq:nonpert_mbmc}) is only true up to $\mathcal{O}(\alpha_\mathrm{QED})$~corrections. We deal with this complication by introducing QED through two ratios~$R$:
\begin{align}
    \frac{\overline{m}_b(\mu)}{\overline{m}_c(\mu)}\Bigg|_{\substack{\mathrm{QCD} \\ \mathrm{QED}}}
    \!\!=\, 
    \frac{R\big(\overline{m}_b/\overline{m}_c, Q_{c,b}=0\!\to\!\tfrac{1}{3}\big)}{R\big(\overline{m}_c(\mu), Q_{c}=\tfrac{1}{3}\!\to\!\tfrac{2}{3}\big)} \times
    \frac{\overline{m}_b}{\overline{m}_c}\Bigg|_{\mathrm{QCD}}.
\end{align} 
Here
\begin{equation}
    R\big(\overline{m}_c(\mu), Q_c=\tfrac{1}{3}\!\to\!\tfrac{2}{3}\big) \equiv\,
    \frac{\mbox{$\overline{m}_c(\mu)$ with $Q_c=\tfrac{2}{3}$}}%
    {\mbox{$\overline{m}_c(\mu)$ with $Q_c=\tfrac{1}{3}$}}
    \label{eq:Rmc}
\end{equation}
is the ratio of the $c$~mass in a theory with $c$-quark charge~$\tfrac{2}{3}$ to the mass in a theory with $c$-quark charge~$\tfrac{1}{3}$. Similarly,
\begin{align}
    R\big(\overline{m}_b/\overline{m}_c, Q_{c,b}=0\!\to\!\tfrac{1}{3}\big) \equiv\,
    \frac{\mbox{$\overline{m}_b/\overline{m}_c$ with $Q_{c,b}=\tfrac{1}{3}$}}%
    {\mbox{$\overline{m}_b/\overline{m}_c$ with $Q_{c,b}=0$}},
\end{align}
where the $c$~and $b$~charges are equal ($Q_c=Q_b$) in each case (and so the ratio is $\mu$~independent).
In every case, the quark masses are tuned to reproduce the continuum meson masses 
in Eqs.~(\ref{eq:mcc_cont}) and~(\ref{eq:mbb_cont}). Either the pseudoscalar or vector mesons can be used; they give the same results to within the precision needed here. Only the first of the $R$~factors (Eq.~(\ref{eq:Rmc})) depends on the $\msb$~renormalization scale~$\mu$; we take $\mu=3$\,GeV, following~\cite{Hatton:2020qhk}. We approximate full QED by \textit{quenched} QED, where only the valence quarks carry electric charge. This is expected to be the dominant contribution in $\mathcal{O}(\alpha_\mathrm{QED})$ and is much less costly to analyze. The techniques we use for introducing QED into simulations are standard and are described in~\cite{Hatton:2020qhk, Hatton:2021dvg}.

\begin{table}
    \caption{Ratio~$R_0$ of $m_{hh}$ with QED corrections to $m_{hh}$ without QED corrections,   evaluated at  the same quark mass~$m_h$. Results are shown for ground-state pseudoscalar and vector mesons analyzed on two configuration sets. The quark's QED charge is $Q$ times the proton's charge; results for $Q=2/3$ can be converted to $Q=1/3$ by replacing $R_0$ with $1 + (R_0-1)/4$.}
    \label{tab:R0}
    \begin{ruledtabular}
    \begin{tabular}{cccll}
    Set & $a \tilde{m}_h$ & $Q$ & $R_0(m_{hh}^P,Q)$ & $R_0(m_{hh}^V,Q)$\\
    \hline
    1 & 0.44935 & $1/3$ & 1.0002907 (26) & 1.0003409 (75)\\
     & 0.59739 & $1/3$ & 1.0002612 (17) & 1.0003106 (30)\\
     & 0.79003 & $1/3$ & 1.0002211 (11) & 1.0002669 (25)\\
    \hline
    3 & 0.27394 & $2/3$ & 1.0015755 (48) & 1.001787 (11)\\
     & 0.39963 & $1/3$ & 1.0003639 (20) & 1.0004081 (55)\\
     & 0.49891 & $1/3$ & 1.0003404 (15) & 1.0003821 (43)\\
     & 0.59739 & $1/3$ & 1.0003182 (14) & 1.0003703 (37)\\
     & 0.69461 & $1/3$ & 1.0002978 (11) & 1.0003543 (37)\\
     & 0.79003 & $1/3$ & 1.00027860 (97) & 1.0003412 (38)\\
    \end{tabular}
    \end{ruledtabular}
\end{table}

\begin{figure}
    \includegraphics[scale=0.9]{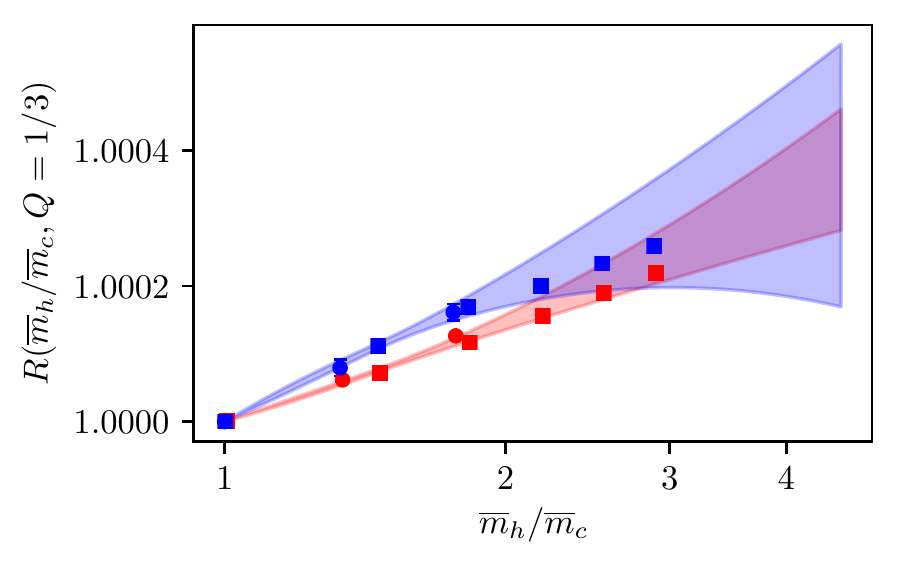}
    \caption{\label{fig:R} Ratio $R(\overline{m}_h/\overline{m}_c,Q=\tfrac{1}{3})$ is plotted 
    versus $\overline{m}_h/\overline{m}_c$. It is the ratio 
    of $\overline{m}_h/\overline{m}_c$ computed with QED charge $Q=\tfrac{1}{3}$ to the 
    result without QED ($Q=0$), where the quark masses are tuned
    to give the same results for $m_{hh}^P$ (bottom, red) or 
    $m_{hh}^V$ (top, blue). Results are shown from configuration
    Sets~1 (squares) and~3 (circles). Errors are smaller than 
    the plot symbols. The blue and red shaded areas show the 
    $\pm 1\sigma$ fits to the data (Eq.~(\ref{eq:Rfit})).}
\end{figure}

The $R$~factor for the $c$~mass (Eq.~(\ref{eq:Rmc})) is the most important and can be inferred from our earlier result~\cite{Hatton:2020qhk}:
\begin{equation}
    R\big(\overline{m}_c(3\,\mathrm{GeV}), Q_c=0\!\to\!\tfrac{2}{3}\big) = \Rmc.
\end{equation}
$R$ is quadratic in~$Q_c$ to better than~0.01\% so the QED correction ($R-1$) required to go from charge~$\tfrac{1}{3}$ to charge~$\tfrac{2}{3}$ is three quarters that required to go from 0~to~$\tfrac{2}{3}$:
\begin{equation}
    R\big(\overline{m}_c(3\,\mathrm{GeV}), Q_c=\tfrac{1}{3}\!\to\!\tfrac{2}{3}\big) = \Rmcred.
    \label{eq:Rmcred}
\end{equation} 

The other $R$~factor, for $\overline{m}_b/\overline{m}_c$, is expected to be much closer to one for two reasons: the QED corrections for the $b$~and $c$~masses are similar and tend to cancel in the ratio; and the charges $Q_{c,b}=\tfrac{1}{3}$ are smaller (and the QED effect is quadratic in the charge). To estimate the effect, we calculated the ratio~$R_0$ of meson masses~$m_{hh}$ with and without $Q_{c,b}=\tfrac{1}{3}$~QED, holding the quark masses constant, for two of our configuration sets; our results are in Table~\ref{tab:R0}. This quantity can be related to the $R$~factor for~$\overline{m}_b/\overline{m}_c$ by re-expressing the $R$-factor in terms of lattice masses, using Eq.~(\ref{eq:nonpert_mbmc}) (since $Q_c=Q_b$), and writing it as
\begin{equation}
    R(\overline{m}_b/\overline{m}_c,Q) = 1 +\frac{\delta\tilde m_b^Q}{\tilde m_b} - \frac{\delta \tilde m_c^Q}{\tilde m_c}
    + \mathcal{O}\big(\delta \tilde m^2\big),
\end{equation}
where $\delta\tilde m^Q_{c,b}$ are the quark mass shifts needed to hold the meson masses constant when QED is added to the simulation. The mass shifts can be calculated for different heavy-quark masses~$m_h$ from the $R_0$ factors in Table~\ref{tab:R0}:
\begin{equation}
    \delta \tilde m_h^Q = \big(1 - R_0(m_{hh}, Q)\big)\,m_{hh}\,\frac{d\tilde m_h}{dm_{hh}}.
\end{equation}
Here the derivative is estimated for each configuration by fitting a cubic spline to the $a\tilde m_{h}$~values in Table~\ref{tab:amhh} as a function of the corresponding $am_{hh}$~values.

Values for $R(\overline{m}_h/\overline{m}_c,Q=\tfrac{1}{3})$ are plotted versus $\overline{m}_h/\overline{m}_c$ in Fig.~\ref{fig:R} for both pseudoscalar (below) and vector (above) mesons from the two configuration sets. We fit these data to a simple function suggested by QED~perturbation theory:
\begin{equation}
    R = 1 + \sum_{i=1}^3 c_i \log^i(\tilde m_h/\tilde m_c) + \sum_{j=1}^5 d_j \big(a\tilde m_h/2\big)^j
\label{eq:Rfit}
\end{equation}
with priors $c_i=0.000(5)$ and $d_j=0.0(5)$. Extrapolating to the $b$~mass gives:
\begin{equation}
    R(\overline{m}_b/\overline{m}_c,Q=\tfrac{1}{3}) =
    \begin{cases}
        \etahRmbmc & \mbox{from $m_{hh}^P$} \\
        \psihRmbmc & \mbox{from $m_{hh}^V$}.
    \end{cases}
    \label{eq:Rmbmc}
\end{equation}
The two results agree with each other, but the corrections are too small to affect our final results significantly.\footnote{$R(\overline{m}_b/\overline{m}_c,Q=\tfrac{1}{3})=1.00059$ to leading order in QED perturbation theory. Our results are close to this value but also include nonperturbative corrections from~QCD.} \red{Doubling the fit priors leaves the results unchanged.} We use the larger error in the error budgets for our final result.

Including both $R$~factors, we arrive at new results for the quark mass ratio at $\mu=3$\,GeV that include (quenched) QED:
\begin{equation}
    \frac{\overline{m}_b(3\,\mathrm{GeV})}{\overline{m}_c(3\,\mathrm{GeV})}\Bigg|_{\substack{\mathrm{QCD} \\ \mathrm{QED}}}
    \!=\,
    \begin{cases}
        \mbmcQetah & \mbox{from $m_{hh}^P$} \\
        \mbmcQpsih & \mbox{from $m_{hh}^V$}.
    \end{cases}
\end{equation}
These again agree with each other. The weighted average, which is our final result, is:
\begin{equation}
    \frac{\overline{m}_b(3\,\mathrm{GeV})}{\overline{m}_c(3\,\mathrm{GeV})}\Bigg|_{\substack{\mathrm{QCD} \\ \mathrm{QED}}}
    \!=\,\mbmcetahQpsihQ.
    \label{eq:mbmcQED}
\end{equation}
The error budgets for  these ratios are the same as those in Table~\ref{tab:errorbudget}, but with an additional uncertainty of 0.03\% associated with the QED correction.\footnote{The QED uncertainty is obtained by adding (in quadrature) the 0.013\% uncertainty in~Eq.~(\ref{eq:Rmcred}), the 0.019\% uncertainty in Eq.~(\ref{eq:Rmbmc}), and 0.017\% for possible corrections due to quenching QED (10\% of the QED correction).} Mass ratios for other values of the renormalization scale are readily calculated using QED perturbation theory,
\begin{align}
    &\frac{\overline{m}_b(\mu)}{\overline{m}_c(\mu)}\Bigg|_{\substack{\mathrm{QCD} \\ \mathrm{QED}}}
    \!\!= \Big(\frac{\mu}{3\,\mathrm{GeV}}\Big)^{{\alpha_\mathrm{QED} }/{ 2\pi}}\,\,
    \frac{\overline{m}_b(3\,\mathrm{GeV})}{\overline{m}_c(3\,\mathrm{GeV})}\Bigg|_{\substack{\mathrm{QCD} \\ \mathrm{QED}}},
    \label{eq:otherscales}
\end{align}
\red{where the  additional QED correction is negligible compared to our errors for typical values of~$\mu$.
Here and elsewhere we ignore the running of $\alpha_\mathrm{QED}$
and $\mathcal{O}(\alpha_\mathrm{QED}\alpha_s)$~corrections since  they are also negligible compared with our errors.}

\begin{figure}
    \includegraphics[scale=0.9]{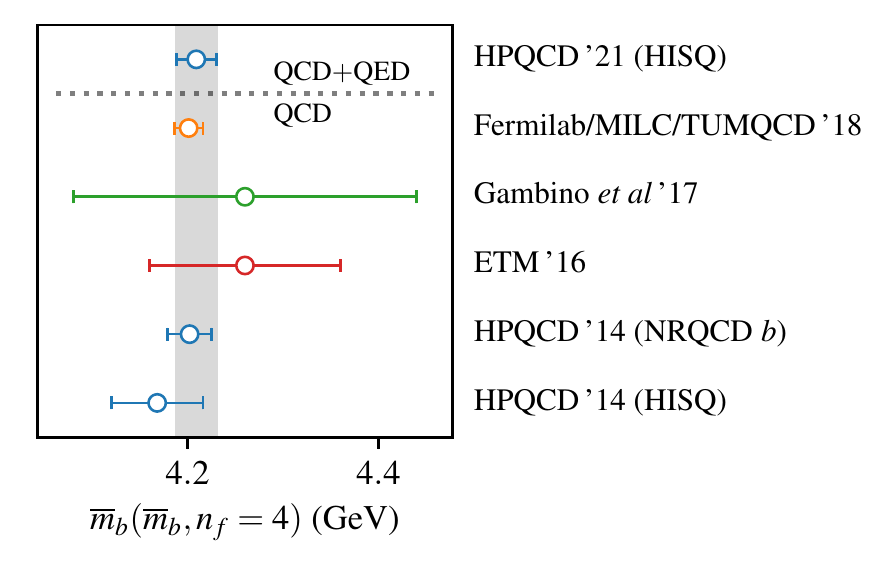}
    \caption{Values for the $\msb$ mass of the $b$~quark from lattice QCD simulations with $n_f=2+1+1$ flavors of sea quark. Results are shown from: HPQCD\,'21 (this paper), Fermilab/MILC/TUMQCD~\cite{Bazavov:2018omf}, Gambino \textit{et al}~\cite{Gambino:2017vkx}, ETM~\cite{Bussone:2016iua}, HPQCD\,'14 (NRQCD)~\cite{Colquhoun:2014ica}, and HPQCD\,'14 (HISQ)~\cite{Chakraborty:2014aca}. The gray band corresponds to the top result (HPQCD~'21), the only one from simulations that include QED.  }
\label{fig:mbm}
\end{figure}

\section{Conclusions}
\label{sec:conclusions}
In this paper we described a new calculation of the ratio of the $\msb$~masses of the $b$~and $c$~quarks:
\begin{equation}
    \frac{\overline{m}_b(3\,\mathrm{GeV},n_f=4)}{\overline{m}_c(3\,\mathrm{GeV}, n_f=4)}\Bigg|_{\substack{\mathrm{QCD} \\ \mathrm{QED}}}
=
    \mbmcetahQpsihQ,
\end{equation}
where $n_f$ is the number of quark flavors in the sea.
This is the first calculation of the mass ratio based on simulations that include QED, and it makes no use of weak-coupling perturbation theory. Earlier analyses used phenomenological models to estimate QED corrections to the mass ratio, but the precision of the most recent results requires a more accurate treatment, like the one described in this paper. QED increased the mass ratio by 0.17(3)\% relative to our ratio without QED (Eq.~(\ref{eq:withoutQED})), which is almost equal to the standard deviation of our final result.\footnote{Note that the ``QED correction'' to a QCD-only analysis depends in detail on how parameters are set in the QCD-only simulation. Since QCD without QED is not the real world, it makes a difference, for example, which hadron mass is used to tune a quark mass; and the QED correction will differ for different choices. Our QED correction is relative to the specific QCD-only theories defined in Section~\ref{sec:withoutQED}.} 

Our new result is consistent at the $1\sigma$~level with earlier results from $n_f=4$~simulations that did not include~QED (and so are $\mu$~independent):
\begin{equation}
    \frac{\overline{m}_b}{\overline{m}_c} = 
    \begin{cases}
        4.578(8)(10) & \text{Fermilab/MILC/TUMQCD~\cite{Bazavov:2018omf}} \\
        4.528(54) & \text{HPQCD~\cite{Chakraborty:2014aca}}.
    \end{cases}
\end{equation}
In both cases the listed uncertainties include estimates of the QED effects.\footnote{The second error in the Fermilab/MILC/TUMQCD result is their estimate of residual QED uncertainties not included in their main analysis (and therefore not included in the errors stated in their abstract)~\cite{Bazavov:2018omf}.} Our new result and HPQCD's previous results are nonperturbative; the Fermilab/MILC/TUMQCD result relies upon perturbation theory (and Heavy Quark Effective Theory), although sensitivity to the perturbative contributions mostly cancels in the ratio. The Fermilab/MILC/TUMQCD result comes from simulations of heavy-light mesons ($D_s$ and $B_s$) rather than the heavy-heavy mesons used here.

In a recent paper,  HPQCD presented a new value for the $c$~mass that includes (quenched) QED effects as we do here:
\begin{equation}
    \overline{m}_c(3\,\mathrm{GeV}, n_f=4)\big|_{\substack{\mathrm{QCD} \\ \mathrm{QED}}}
    = \mcetahQpsihQ\,\mathrm{GeV}.
\end{equation}
Combining this result with our mass ratio gives a new result for the $b$-quark's $\msb$~mass,
\begin{equation}
    \overline{m}_b(3\,\mathrm{GeV}, n_f=4)\big|_{\substack{\mathrm{QCD} \\ \mathrm{QED}}} = \mbetahQpsihQ\,\mathrm{GeV},
\end{equation}
which is the first based on simulations that include QED. Using perturbation theory to run to the $b$~mass gives\footnote{We use 
$\alpha_\msb(5\,\mathrm{GeV},n_f=4)=0.2128(25)$ from~\cite{Chakraborty:2014aca}, together 
with 5-loop results for the beta function and mass anomalous dimension, and 4-loop results 
for adding a flavor~~\cite{Baikov:2016tgj,Herzog:2017ohr,Luthe:2017ttg,Chetyrkin:2017bjc,Baikov:2014qja,Luthe:2016xec,Schroder:2005hy,Kniehl:2006bg,Chetyrkin:2005ia,Liu:2015fxa}.}
\begin{equation}
    \overline{m}_b(\overline{m}_b)\big|_{\substack{\mathrm{QCD} \\ \mathrm{QED}}} = 
    \begin{cases}
       \mbmfouretahQpsihQ\,\mathrm{GeV} & n_f=4 \\ 
       \mbmfiveetahQpsihQ\,\mathrm{GeV} & n_f=5, \\ 
    \end{cases}
\end{equation}
where we now include an evolution factor from QED:
\begin{equation}
    Z_m^\mathrm{QED}(\mu) = \big(\mu/3\,\mathrm{GeV}\big)^{-\alpha_\mathrm{QED}/6\pi}.
\end{equation}
with $\mu=\overline{m}_b$ \red{(which shifts the result by 
less than 0.02\% and so is negligible)}.
This new result for the $b$~quark is compared with earlier results in Fig.~\ref{fig:mbm}. All of these results agree to within errors.

\section*{Appendix}
The $f_{hh}^P(r)$ function plotted in Fig.~\ref{fig:mhmc_vs_mhh} can be recreated from 
the its values at the knot locations $r=\overline{m}_h/\overline{m}_c$,
\begin{equation}
    f_{hh}^P(r) = \left\{\begin{matrix}
        2.9766(13) \\ 4.3951(34) \\ 5.7058(61) \\ 6.9835(93) \\ 8.216(14) \\ 9.435(22)
    \end{matrix}\right.
    \quad \mbox{at $r=$}
    \left\{\begin{matrix}
        1.0 \\ 1.72 \\ 2.44 \\ 3.16 \\ 3.88 \\ 4.6
    \end{matrix}\right. ,
\end{equation}
together with the correlation matrix for these values:
\begin{equation}
    \begin{pmatrix}
        1.0000 & 0.5857 & 0.4282 & 0.3546 & 0.2838 & 0.2113 \\
        0.5857 & 1.0000 & 0.5531 & 0.5205 & 0.4244 & 0.3565 \\
        0.4282 & 0.5531 & 1.0000 & 0.6617 & 0.4088 & 0.4113 \\
        0.3546 & 0.5205 & 0.6617 & 1.0000 & 0.5206 & 0.6074 \\
        0.2838 & 0.4244 & 0.4088 & 0.5206 & 1.0000 & 0.4216 \\
        0.2113 & 0.3565 & 0.4113 & 0.6074 & 0.4216 & 1.0000 \\
    \end{pmatrix}.\nonumber
\end{equation}
The analogous results from the vector mesons are
\begin{equation}
    f_{hh}^V(r) = \left\{\begin{matrix}
        3.09620(20) \\ 4.4836(32) \\ 5.7951(63) \\ 7.054(11) \\ 8.282(16) \\ 9.497(25)
    \end{matrix}\right.
    \quad \mbox{at $r=$}
    \left\{\begin{matrix}
        1.0 \\ 1.72 \\ 2.44 \\ 3.16 \\ 3.88 \\ 4.6
    \end{matrix}\right.
\end{equation}
with correlation matrix:
\begin{equation}
    \begin{pmatrix}
        1.0000 & 0.0999 & 0.0678 & 0.0480 & 0.0403 & 0.0302 \\
        0.0999 & 1.0000 & 0.8547 & 0.6582 & 0.5779 & 0.4747 \\
        0.0678 & 0.8547 & 1.0000 & 0.7891 & 0.6215 & 0.5017 \\
        0.0480 & 0.6582 & 0.7891 & 1.0000 & 0.7001 & 0.5338 \\
        0.0403 & 0.5779 & 0.6215 & 0.7001 & 1.0000 & 0.6628 \\
        0.0302 & 0.4747 & 0.5017 & 0.5338 & 0.6628 & 1.0000 \\
    \end{pmatrix}.\nonumber
\end{equation}
\subsection*{\bf{Acknowledgements}} 
We are grateful to the MILC collaboration for the use of
their configurations. We are also grateful for the use of 
MILC's QCD code. We have modified 
it to generate quenched $U(1)$ gauge fields and incorporate those 
into the quark propagator calculation as described here.
This work used the DiRAC Data Analytic system at the University of Cambridge, operated by the University of Cambridge High Performance Computing Service on behalf of the STFC DiRAC HPC Facility (www.dirac.ac.uk). This equipment was funded by BIS National E-infrastructure capital grant (ST/K001590/1), STFC capital grants ST/H008861/1 and ST/H00887X/1, and STFC DiRAC Operations grant ST/K00333X/1. DiRAC is part of the National E-Infrastructure.
We are grateful to the Cambridge HPC support staff for assistance.
Funding for this work came from the
Science and Technology Facilities Council
and the National Science Foundation.

\bibliography{mbmc}

\end{document}